\documentstyle[11pt,newpasp,twoside]{article}
\def\lya{Ly{$\alpha$}}

\def\etal{et al.}
\def\vrot{$V_{rot}$}
\def\delv{$\Delta v$}
\def\kms{km s$^{-1}$}
\def\lcav{$<$$l_{c}$$>$}
\def\lc{$l_{c}$}

\def\edcomment#1{\iffalse\marginpar{\raggedright\sl#1\/}\else\relax\fi}
\reversemarginpar

\begin{document}
\title{PROBING HIGH-REDSHIFT DISKS WITH DAMPED {\lya} SYSTEMS}
 \author{ARTHUR M. WOLFE}
\affil{University of California, San Diego, CASS, La Jolla, CA 92093-0424}

\begin{abstract}
Evidence is presented that the damped {\lya} absorption systems are the
high-redshift ($z$ $>$ 3) progenitors of galaxy disks.
I discuss kinematic evidence that
the damped {\lya} systems are rotating disks. I also discuss implications of
the lack of metal-poor damped {\lya} systems with 
line width {\delv} $>$ 100 {\kms}. I then present new evidence stemming from
correlations between element-abundance ratios and [Fe/H], which
connects  damped
systems to the thick stellar disk of the Galaxy.   I discuss the connections
between damped {\lya} systems and Lyman break galaxies, and 
how [CII] 158 $\mu$m emission from damped {\lya} systems discriminates
among competing theories of galaxy formation.
\end{abstract}

\section{Introduction}

When did the disks of ordinary galaxies form? This is the central question
to be addressed in my talk. 
At this meeting, evidence was presented 
suggesting the {\em bulk} of the disk population was in place by $z$ $\approx$ 1.
While mergers between disks and infall of gas 
onto disks undoubtedly occur subsequently,
the surveys by Ellis, Lilly, and Faber all rule out substantial changes in
the stellar content of disks since $z$ = 1.
Further evidence against
significant changes in the disk populations comes from the Galaxy. 
From her
studies of stars in the thick disk, Wyse (this volume) concluded that the Galaxy
has not undergone a major merger during the past 12 Gyr.
Provided the Galaxy is typical,
this adds strong support to the idea
that disks, in this case the thick disk, were in place {\em before}
$z$ = 1. 

How long before? This question is crucial since hierarchical cosmologies
predict that only a small fraction of current disk galaxies were in place by
$z$ $\approx$ 3. Rather, CDM models predict the bulk of the protogalactic
mass distribution to be in low-mass subgalactic
progenitors at these redshifts. This is a generic property of CDM cosmogonies,
and one that should be tested. 

In my talk I will describe observations that not only test hierarchical theories
of galaxy formation, but also probe the nature of the protogalactic
mass distribution. The work centers on the damped {\lya} systems, 
because of compelling evidence that these H I layers supply the baryons
comprising visible stars in current galaxies (cf. Storrie-Lombardi
\& Wolfe 2000).
Therefore, 
my talk focuses on gas
rather than stars (but see $\S$ 4).
Gas provides a more accurate picture of the baryonic
content of protogalaxies because most of the baryons at high $z$
were gas.

\section{Kinematics: The Case for Rotation}

\begin{figure}[h]
\vspace{3.6in}
\includegraphics{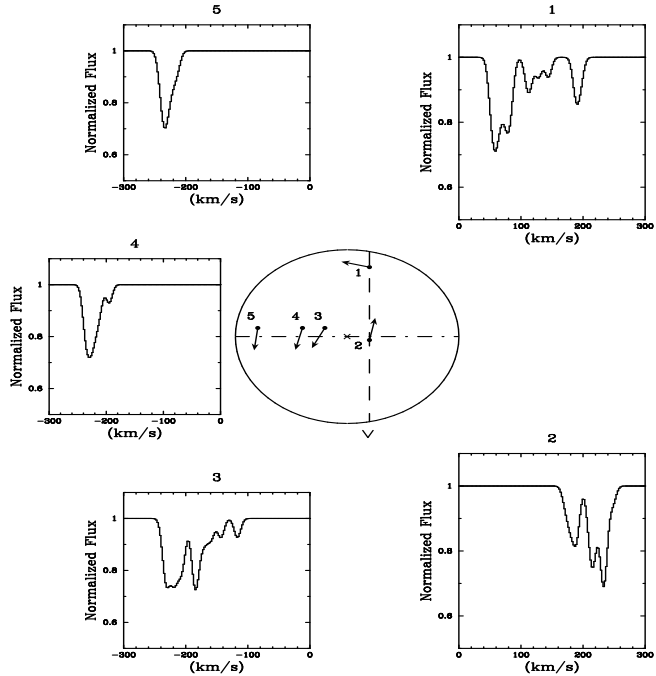}
\caption{``Edge-leading'' velocity profiles produced by
rotating disks (Fig. 3 in Prochaska \& Wolfe 1997). Circle is counter-clockwise
rotating disk with $V_{rot}$ = 250 {\kms}. Dots are intersection points between
sightlines and  midplane of disk. Vertical dashed line is sightline for intersection
points 1 and 2, and horizontal line is kinematic major axis. Absorption profiles
produced at indicated locations} \label{fig-1}
\end{figure}

The evidence for rotating disks stems from accurate velocity
profiles of metal lines that J. X. Prochaska and I have acquired with
HIRES, the echelle spectrograph on the Keck I telescope. Along with
profiles kindly supplied by W. L. W. Sargent, we have acquired accurate
profiles for low-ions such as Fe$^{+}$ and  Si$^{+}$ for over 40 high-$z$
damped {\lya} systems. Low ions are crucial as they
trace the neutral gas which gives rise to damped {\lya} lines.

The velocity profiles (see Wolfe 1999; Prochaska \& Wolfe 1997) consist of
multiple narrow-velocity components  spanning velocity intervals
{\delv} between 20 and 290 {\kms}. The components are not randomly 
distributed in velocity  space. Rather, the component with the largest
optical depth is at the profile edge in 75 $\%$ of the profiles with
{\delv} $>$ 40 {\kms}.
The most straightforward explanation for this ``edge leading'' asymmetry
is that the damped {\lya} sightlines traverse rotating disks in which the
density of absorbing clouds decreases exponentially in the radial and vertical
directions with scale lengths $R_{d}$ and $h$ (Prochaska \& Wolfe 1997).
Figure 1 shows how ``edge-leading'' profiles
naturally arise from randomly selected impact parameters. The figure also
shows why {\delv} decreases with increasing impact parameter. 
While other mechanisms
have been suggested to explain the profile asymmetries (e.g. Haehnelt {\etal}
1997), exponential
rotating disks  provide the most straightforward
explanation.

Using 
Monte Carlo techniques,  we tested
the semi-analytic models of Mo {\etal}  (1998) in which  
centrifugally-supported disks are at the centers  of dark-matter halos with mass distribution
determined by Press-Schecter functions and CDM power spectra. We generated
synthetic velocity profiles by letting randomly selected sightlines traverse 
randomly-oriented exponential disks with rotation speeds {\vrot} equal
to $\sqrt{GM(r)/r}$ at the virial radius.
We found the {\delv} predicted by single-disk CDM models cannot account
for the significant fraction of {\delv} between 100 and 300
{\kms}, since the median {\vrot} predicted by 
single-disk CDM models are less than 100 {\kms}.
However, the TF model in which the input distribution of {\vrot}
at high redshift is determined by the Tully-Fisher relation and
the Schecter luminosity function
is compatible with the data owing to the larger fraction of disks {\vrot}
$>$ 100 {\kms} (Wolfe \& Prochaska 2000).
Thick disks with $h$$\approx$ 0.3$R_{d}$
are required to obtain {\delv} $>$ 100 {\kms}.

\section{Metals in Damped {\lya} Systems}

\begin{figure}
\vspace{2.5in}
\includegraphics{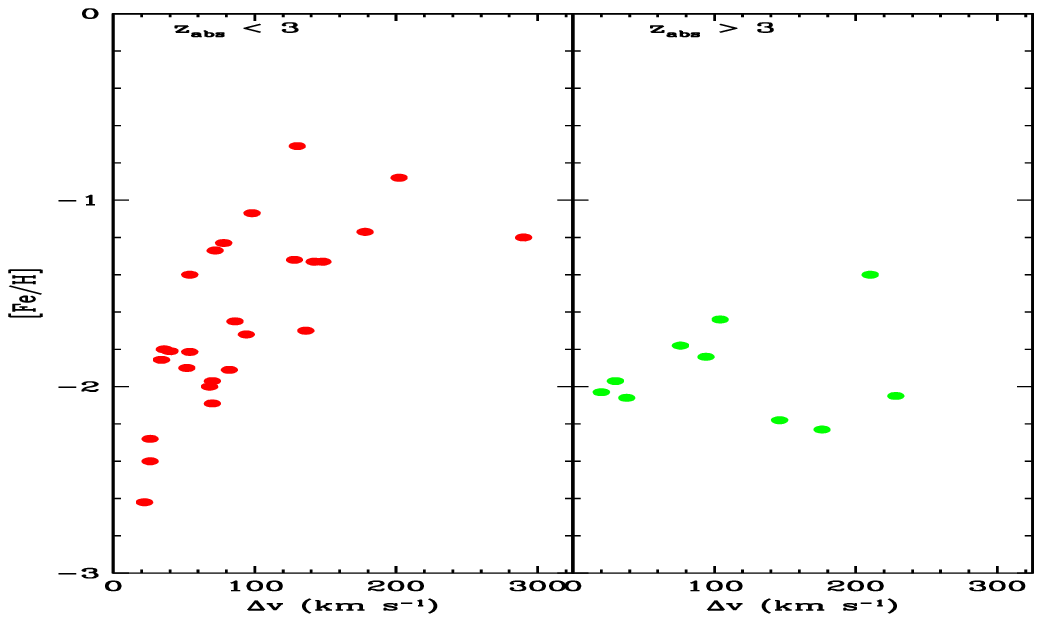}
\caption{[Fe/H]  versus {\delv} for damped {\lya} systems at
(a) $z$ $<$ 3 and (b) $z$ $>$ 3, where [Fe/H]=log(Fe/H)-log(Fe/H)$_{\odot}$} \label{fig-2}   
\end{figure}

\subsection{Metallicity versus Kinematics in Damped {\lya} Systems}

We also used
the HIRES velocity profiles 
to determine the metallicities of damped {\lya} systems
(Prochaska \& Wolfe 2000). Combining  metallicity and kinematic
information we find a phenomenon that bears upon the presence 
of rotating disks at high redshifts.
And this is the clear deficit at $z$ $<$ 3  of metal-poor systems with  {\delv} $>$
100 {\kms} shown in Figure 2a. The absence of these systems is real and not an artifact due
to selection effects. This is indicated by the presence of  such systems in the
$z$ $>$ 3 sample (Fig. 2b). Moreover, dust would lead to the
absence of metal-rich rather than metal-poor systems. The enlarged
sample of [Fe/H] abundances strengthens the case 
first made with a smaller sample of  
[Zn/H], {\delv} pairs (Wolfe \& Prochaska 1998). We interpret this effect to arise
from negative radial gradients in [Fe/H] in rotating disks. High [Fe/H] are
required at small radii 
because low
impact parameters, $b$, are a necessary condition for large {\delv} (see Fig. 1).
We find a wider distribution of $b$ contributes to systems with
smaller {\delv}. Thus, at small {\delv}, the lower [Fe/H] come from  
low [Fe/H]  at large radii, while higher [Fe/H] again come from small $b$ 
but in disks that  are more face-on. {\em Therefore, the increase in scatter of  [Fe/H]
with decreasing {\delv} is naturally produced by the rotating disk models.} 
However,
since variations in metallicity in nearby spirals rarely span more than 1 dex,
a positive correlation between [Fe/H] and $M$ would help to explain
the full effect.

\subsection{Abundance Patterns and the Thick Disk Connection}

\begin{figure}
\vspace{4.0in}
\includegraphics{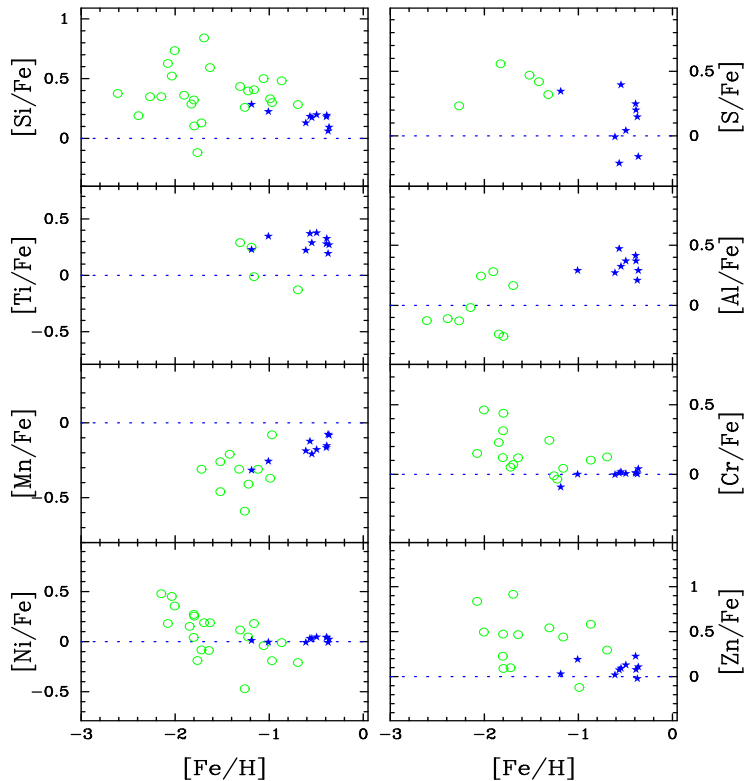}
\caption{Abundance ratios versus [Fe/H]  for damped systems (circles)
and thick disk stars (stars) (from Prochaska {\etal} 2000)} \label{fig-3} 
\end{figure}

Figure 3 plots abundance ratios of various elements versus [Fe/H] for damped {\lya} systems and 
stars in the Galaxy thick disk. Comparison with the thick disk is motivated by 
the large vertical scale-heights required in our model gaseous disks, since 
this raises the possibility
that high-$z$ damped {\lya} systems are the progenitors of thick stellar
disks. To demonstrate the plausibility of this hypothesis, we showed
that the metal-rich damped {\lya} systems contain sufficient baryons
at mean metallicity,  
$<[{\rm Fe/H}]>$ = $-$ 0.6, 
to account for the 
the mass content and metallicity of thick disks (Wolfe \& Prochaska 1998).
Here I discuss
further evidence in support of this idea.

The abundances in Figure 3 were inferred from HIRES data obtained for a kinematically
selected sample of stars in the thick disk and our sample of damped {\lya}
systems. The stellar atmospheres programs used to deduce the thick disk abundances are
described in Prochaska {\etal} (2000). The figure plots results for $\alpha$ , Fe peak, and light
elements. As [Fe/H] increases we find tentative evidence that [X/Fe] for the damped systems evolves
toward [X/Fe] for the thick disk when X = Si, Zn, and Mn. The results are especially striking for
[Mn/Fe]. We find two possible interpretations for these trends. The first is a nucleosynthetic
explanation. Lu {\etal} (1996) argued that the damped systems exhibit classic
type II Sn patterns with enhanced [$\alpha$/Fe] ratios, deficient [Mn/Fe] ratios
(due to the odd-even effect), and unpeculiar [Cr/Fe] and [Ni/Fe] ratios. The
problem with this argument was that [Zn/Fe] = 0 in stars with a wide range in
metallicities; whereas [Zn/Fe] $\approx$ 0.3 in damped systems. With our more accurate
spectroscopy we find [Zn/Fe] = 0.1 in thick-disk stars where [Fe/H] = $-$ 0.6.
And there is tentative evidence that [Zn/Fe] is higher for thick-disk stars with
lower [Fe/H]. That Zn does not track Fe is not
surprising since recent calculations indicate Zn is not produced
in the same type I Sn responsible for most of the Fe. Rather, Zn may
be produced in neutrino-driven winds in type II Sn (Hoffman 1996)).
Therefore, it is likely that the [Zn/Fe] ratio in damped {\lya} systems
is partially nucleosynthetic in origin.
On the other hand, Pettini {\etal} (1997) argued that the abundance patterns
in damped systems are caused by depletion of heavy elements on to dust grains. This is
consistent with all the observed patterns except for [Mn/Fe] and possibly [Ti/Fe].
The strongest argument in favor of this explanation was the high [Zn/Fe] ratio. But
we have shown that
part of this ratio probably originates from nucleosynthesis, and so a more likely 
explanation is that the damped {\lya} abundance pattern 
results from small amount of dust superposed on a type II Sn 
abundance pattern.

As a result, we conjecture that metal-poor damped {\lya} systems evolve into metal-rich
objects that become thick stellar disks at high redshifts, say $z$ $\approx$ 3.
Since the velocity dispersion of the model gaseous disks, $\sigma$ $\approx$
10 {\kms}, subsequent merger events are required to heat the stars to the
observed velocity dispersion of the thick disk, $\sigma$ $\approx$ 40 {\kms}.
In that sense our model does not differ substantially from the scenario put
forward by Wyse (these proceedings).

\section{Connecting Lyman Break Galaxies to Damped {\lya} Systems}

For the past two  years, our group (Gawiser, Prochaska, Cooke, \& Wolfe) has 
been searching for Lyman break galaxies in fields centered on quasars
with foreground damped {\lya} systems. Our goals are to (i) find galaxies physically
associated with the damped systems, and (ii) determine  how often the
damped {\lya} redshifts coincide with the redshift spikes characterizing the distribution
of known Lyman break galaxies (Steidel 1999).
The correlation between damped systems and Lyman break galaxies is very interesting
because it tells us about the bias of the damped {\lya} galaxies, and, as a result,
about their mass. Stated differently, 
if damped {\lya} redshifts are preferentially located
in Lyman break redshift spikes, this would mean the 
amplitudes of density
fluctuations,
$(\delta \rho/\rho)_{DLA}$,
destined to become
damped {\lya} galaxies are too low to collapse and form galaxies
unless they are boosted by the enhanced mass density provided by
the Lyman break clusters. This would indicate higher-than-typical masses
given the nature of the power spectrum.

Our survey initially focused on $B$ drop outs as there is no $U$ band photometry
currently available on the Keck telescopes. This restricted the sample to damped
{\lya} systems with $z$ =(3.8,4.5). We have acquired  $B$,$R$,$I$ images for
about four damped {\lya} fields in which the 1$\sigma$ sensitivies correspond to
to $B$ $\approx$ 29.5, $R$ $\approx$  26.5, and $I$ $\approx$ 26.0. 
As a result we are able to detect
blue drop-outs down to $R$ $\approx$ 26.0. We select candidate blue drop-outs
using criteria established by Steidel and then acquire multi-object spectra
for about 20 candidates per setting using the LRIS multi-object spectrograph.
The data have been reduced for one field containing damped systems at $z$ = 3.871 and
4.072. While there are no confirmed galaxies near the high redshift systems,
we have identified four galaxies within $\Delta z$ = 0.02 of the lower redshift
system. That is, four out of a total of twelve identified galaxies lie within a Steidel-like
redshift spike that includes the damped redshift. If future observations show this
this effect to be generic, then some fraction of damped systems may evolve into
galaxies likely to be found in cluster environments, such as ellipticals.

\section{[C II] 158 $\mu$m Emission from Damped {\lya} Systems}

[C II] 158 {\micron} emission results from transitions between
the P$^{2}_{3/2}$ and P$^{2}_{1/2}$ fine structure states in 
C$^{+}$.
It dominates cooling by the Galaxy ISM with a luminosity $L$([C II])
=5$\times$10$^{7}$$L_{\odot}$ (Wright {\etal} 1991).  
Significantly, most of the 158 {\micron}  emission from the ISM in
the Galaxy
and in other spiral galaxies  where it  is detected 
comes from the cold neutral medium (CNM) rather than the warm 
neutral medium, 
star-forming regions in spiral arms (Madden {\etal} 1993), or PDRs on
the surfaces of molecular clouds.
The last point is especially
relevant for damped {\lya} systems where molecules are rarely
detected (cf. Lu {\etal} 1999).

The 158 {\micron} luminosities of
damped {\lya} systems can be estimated from $N$(C II$^{*}$), the column density
of C$^{+}$ ions in the P$^{2}_{3/2}$ state, which
is inferred from the absorption profiles of the UV
transition C II$^{*}$ 1335.708. The 158 {\micron} luminosity
per H atom is given by

\begin{equation}
l_{c} = h{{\nu}_{21}}N({\rm C II}^{*}){\rm A}_{21}/N({\rm H I}) {\rm erg s^{-1} H}^{-1},
\end{equation}

\noindent where $h{{\nu}_{21}}$ is the energy of the transition and $A_{21}$ 
is the coefficient for spontaneous photon emission (Lu {\etal} 1999). 
Since $L$([C II])=$<$$l_{c}$$>$$M$(H I)/$m_{H}$, where $<$$l_{c}$$>$ is the
density-weighted average of $l_{c}$, we can estimate the C II luminosity
from damped systems provided we know the H I mass and $<$$l_{c}$$>$.

In principle we can determine {\lcav} from the distribution of
{\lc} measured along individual sightlines. 
Figure 4 shows estimates of {\lc}
for 11 
damped {\lya} systems. 
We also show {\lc} deduced
for separate sightlines through the ISM and {\lcav}
estimated for the ISM. We compute {\lcav} for the damped sample 
by assuming  exponential disks. In that case  {\lcav}
$\approx$ {\lc} for impact parameters $b$ $\approx$ 2$R_{d}$. 
Our Monte Carlo simulations show that most damped sightlines have
$b$ $<$ (4$-$5$)\times$$R_{d}$. As a result  
{\lcav} corresponds to largest (2/4.5)$^{2}$$\times$11
$\approx$ 2$-$3 values of {\lc} in Figure 4.
We estimate log{\lcav} = $-$26.6  $\pm$0.2. While somewhat
uncertain, {\lcav} for the damped systems is
clearly more than 30 times lower than for the ISM. 
Since
the [C II] 158 {\micron}
line is the dominant coolant for the CNM, 
{\lcav}
equals the heating rate for steady state conditions.
As a result, the low value of {\lcav} indicated for damped 
{\lya} systems suggests low heat inputs into the protogalactic gas
at high redshifts.

\begin{figure}
\vspace{3.3in}
\includegraphics{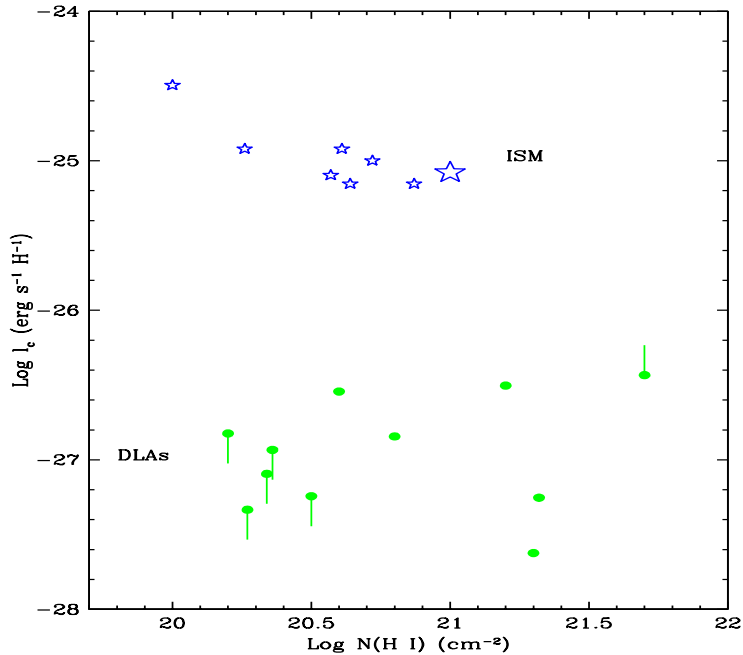}
\caption{[CII] 158 {\micron} emission per H atom versus N(HI). 
Circles are
damped {\lya} systems with upper limits and one lower limit shown. 
Small stars are estimates for separate ISM sightlines. 
Large star is global average of ISM from COBE}\label{fig-4} 
\end{figure}

We also use {\lcav} to compute the 
distribution of $L$([C II]) 
of damped {\lya} systems in various
cosmogonies. Mo {\etal} (1998) predict
the fraction of intercepted damped {\lya} systems with 
dark-halo masses exeeding $M$. We 
convert $M$ to 158 {\micron} emission noting that 
$L$([C II])={\lcav}$m_{d}$$M$/${\mu}m_{H}$, where $m_{d}$ is the fraction
of dark halo mass in the disk. 
Assuming log{\lcav}=$-$26.5
erg s$^{-1}$H$^{-1}$,
$\mu$ = 1.4, 
and 
$m_{d}$=0.05, we find $L$([C II])=2$\times$10$^{7}$($M/10^{12}M_{\odot}$)$L_{\odot}$.
We
compute the flux density at a given redshift by assuming a rectangular 
velocity profile with width = 2$\times$$V_{rot}$sin($i$), where sin($i$) = 0.5.
The results for the ${\Lambda}$CDM and TF cosmogonies are shown in 
Figure 5. At $z$ $\approx$ 3
the predicted ALMA
detection rate is $<$ 3 $\%$ the CDM model, 
and 
about 25 $\%$ in the TF model. Thus detection of [C II] 158 {\micron}
emission in high-$z$ damped systems would be highly significant. Moreover,
measurements of the velocity profile of the emitting gas could distinguish
rotating disks from protogalactic clumps (Haehnelt {\etal} 1997).

\begin{figure}
\vspace{2.8in}
\includegraphics{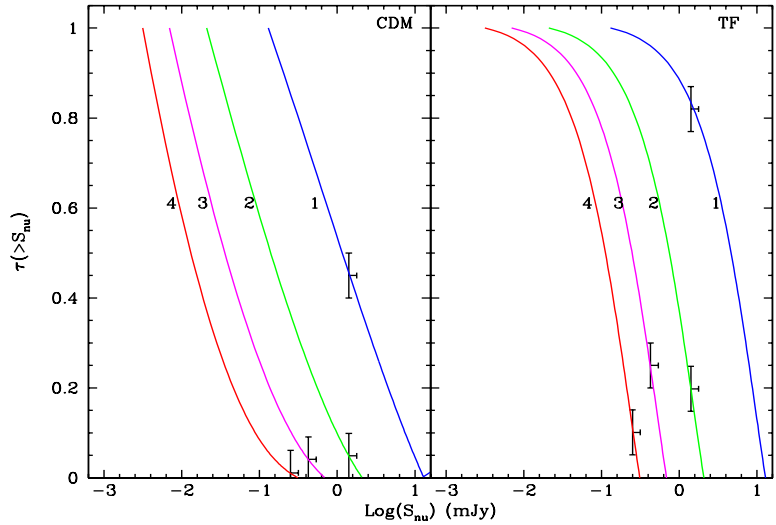}
\caption{Fraction of damped {\lya} systems
detected with 158 {\micron} flux density exceeding
$S_{\nu}$. (a) Curves are predictions for ${\Lambda}$CDM model
at redshift $z$ = 1,2,3,4. Error bars are 3$\sigma$ ALMA sensitivities 
for S$_{\nu}$  at each redshift (Brown 2000).
(b) Same as (a) except for TF model}\label{fig-5}
\end{figure}

\begin{quote}
\verb"acknowledgments"
I wish to thank my collaborators E. Gawiser and  J. X. Prochaska for permission
to quote our joint research. This work was partially supported by NSF grant
AST0071257.
\end{quote}

\end{document}